# Charge-transport and tunneling in single-walled carbon nanotube bundles

M. Salvato[a,b], M. Cirillo[a], M. Lucci[a], S. Orlanducci[c], I. Ottaviani[a], M. L. Terranova[c], and F. Toschi[c]

[a] *Dipartimento di Fisica and MINAS Laboratory, Università di Roma "Tor Vergata", I- 00133 Roma*

[b] *Laboratorio Regionale "Super Mat" CNR-INFM, I-84081, Baronissi, Italy*

[c] *Dipartimento di Scienze e Tecnologie Chimiche and MINAS Laboratory, Università di Roma "Tor Vergata", I- 00133 Roma*

**Abstract**

We investigate experimentally the transport properties of single-walled carbon nanotube bundles as a function of temperature and applied current over broad intervals of these variables. The analysis is performed on arrays of nanotube bundles whose axes are aligned along the direction of the externally supplied bias current. The data are found consistent with a charge transport model governed by the tunnelling between metallic regions occurring through potential barriers generated by nanotube's contact areas or bundles surfaces. Based on this model and on experimental data we describe quantitatively the dependencies of the height of these barriers upon bias current and temperature.

Carbon nanotubes (CNT) have attracted interest in the past decade both at fundamental and applied physics level due to the intriguing nature of the excitations at the basis of the conduction processes [1,2] and to the appealing potentiality in electronic applications [3]. Although work has been recently focused on isolated CNT [2,3], the transport properties of CNT aggregates also attracts interest because of their applications as thin film transistors and nanovalves [4], chemical sensors [5], and conducting fillers in otherwise insulating materials [6].

In spite of the large effort devoted to the understanding of the charge transport mechanism in aggregates, a comprehensive and consistent picture of this process has not yet been achieved. The Variable Range Hopping (VRH) [7] and the thermal activation model [8,9] have received interest, but their predictions rarely allow fittings of experimental data in a wide range of temperature. In particular, in the limit of low temperature the resistivity of CNT assumes a finite value rather than diverging [6,9,10], a behavior observed even for semimetallic materials and doped polymers [11,12]. Among the theoretical models that have been proposed to explain the observed experimental features [6,9-12] the Fluctuation Induced Tunneling (FIT) model [13] has also been subject of attention [11]; in this model the conduction process is attributed to tunneling between normal metal portions of the CNT. In this letter we present a systematic experimental analysis showing that, for single–walled (SW) CNT bundles aligned along the bias current direction, the FIT model can account for the experimental observations over broad temperature and current ranges.

We deposited SWCNT on $SiO_2$ substrates on which metallic thin films had been previously patterned in order to provide the electrical contact. The electrodes, consisting of Au, Al or NbN, had multifinger shape consisting of 20 μm spaced parallel stripes alternately connected to two electrodes as shown in the optical microscope photo of Fig. 1a (picture on the left). The CNT were aligned along the direction ortogonal to the fingers by a dielectrophoretic technique described elsewhere [5]. A scanning electron microscope (SEM) picture of aligned SWCNT bundles bridging two fingers (follow the arrows in the photo) is shown in Fig. 1b. Pairs of Cu wires were soldered on the two metallic electrodes and connected to a data acquisition system as sketched in the rightmost picture of Fig. 1a. The samples were stuck with metallic paste over the cold finger of a high vacuum cryocooler which allowed to change the sample temperature in the range (5-300)K. The experimental data do not show any dependence on the electrode metal even in the case of the superconducting NbN ; in this case, below the superconducting critical temperature ($T_c$ = 14 K) no contribution to the sample resistance is given by the multifinger.

Fig.2a shows the normalized resistance vs. temperature measurements at different bias currents for SWCNT contacted with Al electrodes ; for each curve the resistances are normalized to their values $R_0 \approx 1.2$ kΩ measured at 250 K. All the samples show a semiconducting character with a metallic component evidenced by the finite value assumed by the electrical resistance in the limit of zero temperature. We see that the influence of the bias current on the electrical resistance is very effective and in particular an increasingly metallic behavior is clearly observed at higher bias currents. The inset of Fig. 2a displays some of the curves of this figure in a semilogarithmic plot where the inverse of the temperature is on the horizontal axis; as we shall see this kind of plot turns out to be very convenient for the purposes of our analysis. Although non linear resistance is expected in the case of localized systems, the VRH model does not provide a fit to our data : plotting the logarithm of the resistivity as a function of $(1/T)^\gamma$ with $\gamma = 1/2$, 1/3 or 1/4, we obtain non linear dependencies.

The possible influence of a contact resistances between electrodes and CNT bundles was ruled out by a four contacts geometry measurement; the results of this test are reported in Fig. 2b for four values of the bias current. Comparing these data with those of Fig. 2a (inset) we see that the shapes of the resistivity curves are very similar independently upon the contact geometry. We conclude that the contact resistance present in two leads contacted samples adds linearly to the CNT resistance and that possible Schottky barriers formed at the CNT-electrode interface have a negligible effect on our measurements.

In Fig.3a we report dependencies of the resistivity for different currents as a function of $1/T$ in a semilogarithmic plot for the NbN contacted samples. Apart for the jumps at $T$=14 K (corresponding to superconducting transition of the NbN film), the shape of the curves is similar to that obtained for CNT deposited on Al electrodes shown in Fig. 2. However, below 14 K we have no contribution of the NbN electrodes to the resistance. At high temperature the linear $1/T$ dependence of the electrical resistance, expected by the Arrhenius law, suggests that thermal activation governs the transport mechanism. At low temperature, instead, both VRH and thermal activation [7,8] predict a divergence in the resistance which is clearly not our case. Moreover, we note that for Coulomb blockade based model in isolated CNT a non linear dependence of the resistance as a function of $1/T^\alpha$ is predicted [14]: this functional dependency clearly does not fit the data of Fig. 2 and Fig. 3a.

Let us analyze the data of Fig. 2 and 3a within the framework of the FIT model [13]. Assuming a parabolic spatial shape of the barrier energy between the conducting regions this model predicts the following exponential dependence of the resistance upon the temperature [13]:

$$R = R_0 e^{T_1/(T+T_0)} \tag{1}$$

in this expression $R_0$ is the resistance at room temperature, $T_1 = 2SV_0^2 / \pi k_B e^2 w$ and $T_0 = 4\hbar S V_0^{3/2} / \pi^2 w^2 k_B e^2 \sqrt{2m}$ with $S$ and $w$ being the junction surface and width respectively, $V_0$ is the depth of the potential well, $m$ the electron mass, $e$ the electron charge, and $k_B$ and $h$ ($\hbar = h/2\pi$) are respectively the Boltzmann and Planck constants. In the case of a network of CNT aligned along the bias current direction, one can suppose that the charge transport occurs mainly along the metallic tubes surface and that electron tunneling takes place across the connection between CNT and bundles. These connections can be considered as junctions of insulating defects along an otherwise metallic path and their presence gives rise to the distortion of the energy levels and a change of the density of states at the boundaries [1]. $V_0$ can be interpreted as the potential barrier that the electrons have to overcome in order to tunnel through these defects and depends on the electronic structures present on each side. In semiconductors [15] and CNT [16] tunnelling, originated by structurally stable barriers, the external driving forces and thermodynamic parameters influence the relative electronic density of states across the barrier and the potential energy associated to the barrier is assumed to depend on the bias current and temperature [15,16,17]. The equivalence of such physical effects in our case is the dependence of the height of the barrier $V_0$ upon temperature and bias current.

The solid curves in Fig.2 and Fig. 3a are fits to the data based on eq. (1). For the Al contacted samples, the fit is very good over the whole temperature range suggesting that FIT mechanism can be invoked for these aligned CNT. Also, the values of the fitting parameters $T_1/T_0$ reported in the captions of the figures, are consistent with previous investigations.

The linear extrapolation of the normalized resistance curves in Fig. 3a at zero temperature provides the value of the ratio $T_1/T_0 = \pi w \sqrt{2mV_0}/2\hbar$ for four different currents and, as a consequence, the depedence of $V_0$ as a function of the bias current. In Fig.3b indeed we report the result that we have obtained. We see that at low current bias levels ($I < 1$ μA) a deviation toward a constant energy barrier regime is evident whereas at high current regime the data are fitted by the expression $wV_0^{1/2} = -A \log I$ with $A = 0.12$ nm·eV$^{1/2}$. The "saturation" value for very low currents indicates that in this limit a substantial modification of the barrier cannot be generated by the current, a result which is reasonable. The data of Fig.3b allow to estimate the maximum value of $V_0$ = 1.2 eV if a van der Waals distance $w = 0.34$ nm is assumed between the metallic parts of the tunnel junction [18].

The value of 1.2 eV found above is consistent with measurements reported in [6] referring to nanotube polymers composite and interpreted on the basis of the FIT model. In particular, the authors of ref. 6 found values of the FIT parameters $T_1$ and $T_0$ which correspond to values of the potential barriers in the range 0.45 eV - 1 eV when the nanotube concentration inside the polymer increases from 8% to 25%. This indicates that $V_0$ increases when a large number of contact between CNT or bundles is made suggesting that tunnel is dominant between the different bundles and negligible inside each bundle. Our estimate for $V_0$ allows to exclude the possibility to attribute our results to resistive behavior given by Coulomb blockade which is present in this kind of structures due to Luttinger liquid electron transport models [14]. This phenomena is in fact related to an increase of the barrier voltage due to the addition of single electrons that tunnel the barriers of the order of few meV. These low values of the barrier height are usually observed when the electrodes are single CNT and the barrier is the contact between them. In our case, the barrier $V_0$ refers to the contact between different bundles and it is about three order of magnitude higher as follows by our experimental data.

Similarly to other physical situations [15,17], one can suppose that this potential barrier $V_0$ can depend on the temperature and on the driving force which favors the migration of the charges between two adjacent metallic regions. We will show in what follows that, from an hypotesis of functional dependencies of the potential barrier upon current and temperature, complete consistency with our experimental results can be found. The dependence of $V_0$ upon temperature cannot be obtained by the slope of the $R$ vs. $T$ curves since both $T_1$ and $T_0$ in eq. (1) depend on $V_0$. This dependence can be determined, however, from the *I-V* characteristics. We assume at this point that the potential energy depends on both temperature and current in a way that the function $V_0(T, I)$ is separable in these two variables. The general expression for $V_0(T, I)$ can be written as follows:

$$V_0(T,I) = V_I(T)\ln^2(I) \tag{2}$$

where $V_I(T)$ takes into account for the temperature dependence of $V_0$ and in the rightmost term we have indeed substituted the analytical form suggested by the fit of the data in the bias region above 1 µA of Fig. 3b, namely $V_0 \propto \ln^2(I)$. Replacing $V_0(T,I)$ given by eq. (2) in the expressions of $T_0$ and $T_1$ and inserting these inside eq. (1), one obtains the following expression for the normalized resistance as a function of both temperature and current in the limit of high bias currents:

$$\ln\frac{R}{R_0} = \alpha V_I(T)^{1/2} \ln(I) \tag{3}$$

where $\alpha = 0.4$ eV$^{-1/2}$. Fig. 4a shows double logarithmic plots of the current dependence of the electrical resistance, as obtained by the *I-V* measurements, at different temperature for our SWCNT. For higher bias currents the logarithmic dependence is in very good agreement with eq. 3, as shown by the straight lines fit to the data. For low currents and low temperatures the conduction is likely regulated by the tunneling of electrons through the barriers generating a constant value (for each given temperature) of the resistances. For low currents and high temperatures instead thermally excited electrons overcome the lowered energy barriers and are likely responsible for the conduction process: for each temperature the thermal excitations provide a given number of electrons and therefore their value does not depend upon the currents. We believe the two just described processes contribute to generate the ohmic tendencies that we observe in Fig. 4a ; in any case our data follow the predictions of eq. (3) in the limit of high bias current where our approximation is supposed to provide a better fit to the data.

Following eq. (3), the temperature dependence of the potential barrier $V_I(T)$ is given directly by the logarithmic slope of the curves at high bias currents where our approximation clearly works. Fitting the linear part of the data in Fig.4a the values of the fit parameters found are between 0.37 and 0.12 giving values for $V_I(T)$ in the range 0.9 - 0.1 eV for all the data measured between $T = 8$ K and $T=280$ K respectively. Fig.4b shows the $V_I(T)$ (normalized to its value at 280 K) vs. $T$ dependence. The decrease of the potential barrier due to thermal effects is clear for all the samples and follows an exponential law dependence $e^{-T/T_{0C}}$ with $T_{0C} = 277$ K. This exponential behavior is close to dependencies of barrier height on temperature observed in other physical systems [17] and therefore we find it reasonable.

In conclusion we have explained systematic experimental results on charge transport in SWCNT bundles assuming a tunnel mechanism between the metallic regions separated by insulating barriers and enhanced by thermal fluctuations. Due to the particular geometry obtained by the deposition process, the presence of the investigated potential barriers is attributed to the contact between the different bundles forming the chain between two electrodes. Our results indicate that the transport mechanism cannot be explained as simple scattering of electrons via impurities and phonons as in the case of metallic samples and that it can be consistently interpreted in terms of the fluctuations induced tunneling (FIT) model.

# Figure captions

**Figure 1.** a) Optical image and sketch of the biasing condition for our multifinger configuration. For this particular sample the spacing between the fingers was 10 μm; b) SEM image showing (follow the arrows) bundles of SW CNT connecting two fingers. The current fed to the multifingers flows along the axis of the bundle. The image has been obtained by merging 4 SEM photos. The upper photo shows a general view of the bundles organization/alignement at the contact electrodes. The contact electrodes are indicated in the figure (lighter gray areas).

**Figure 2.** a) Resistance, normalized to their value $R_0$ measured at T=250K, vs. temperature for SWCNT deposited on Al electrodes dependencies at different bias currents (top to bottom : 0.1 μA, 0.5 μA, 5μA, 10μA, 50μA, 100μA, 500 μA, 1mA). The inset shows the data in a $R$ vs. $1/T$ plot. The lines are fits to the data obtained from eq. (1) with fit parameters: $T_1/T_0$=26.67, 7.37, 2.88, 1.74 and $T_0$=9.1K, 38K, 136K, 381K for $I$= 1 μA to 500 μA respectively; b) four contact measurement performed for comparison with a) ; lines are fit to the data through eq. 1 with fit parameters: $T_1/T_0$ =55, 25, 7.8, 5.2, and $T_0$=3.16, 6.7, 30, 55 for $I$= 1 μA to 500 μA respectively.

**Figure 3**. a) Resistance vs. Temperature plot at different bias currents for SWCNT deposited on NbN electrodes. The lines are fits to the data obtained from eq. (1) with fit parameters: $T_1/T_0$=23.5, 12.0, 3.8, 1.75 and $T_0$=17.4 K, 35.1 K, 138 K, 569 K for $I$= 3 μA to 1 mA respectively; b) square root of the potential barrier $V_0$ as a function of the bias current in the limit of zero temperature for SWCNT. The parameter $w$ on the vertical axis label represents the width of the barrier. The current axis is logarithmic.

**Figure 4.** a) Log-log plot obtained from measurements of current-voltage characteristics at different temperatures ; b) Normalized potential energy as a function of the temperature as obtained from the straight lines fittings in a).

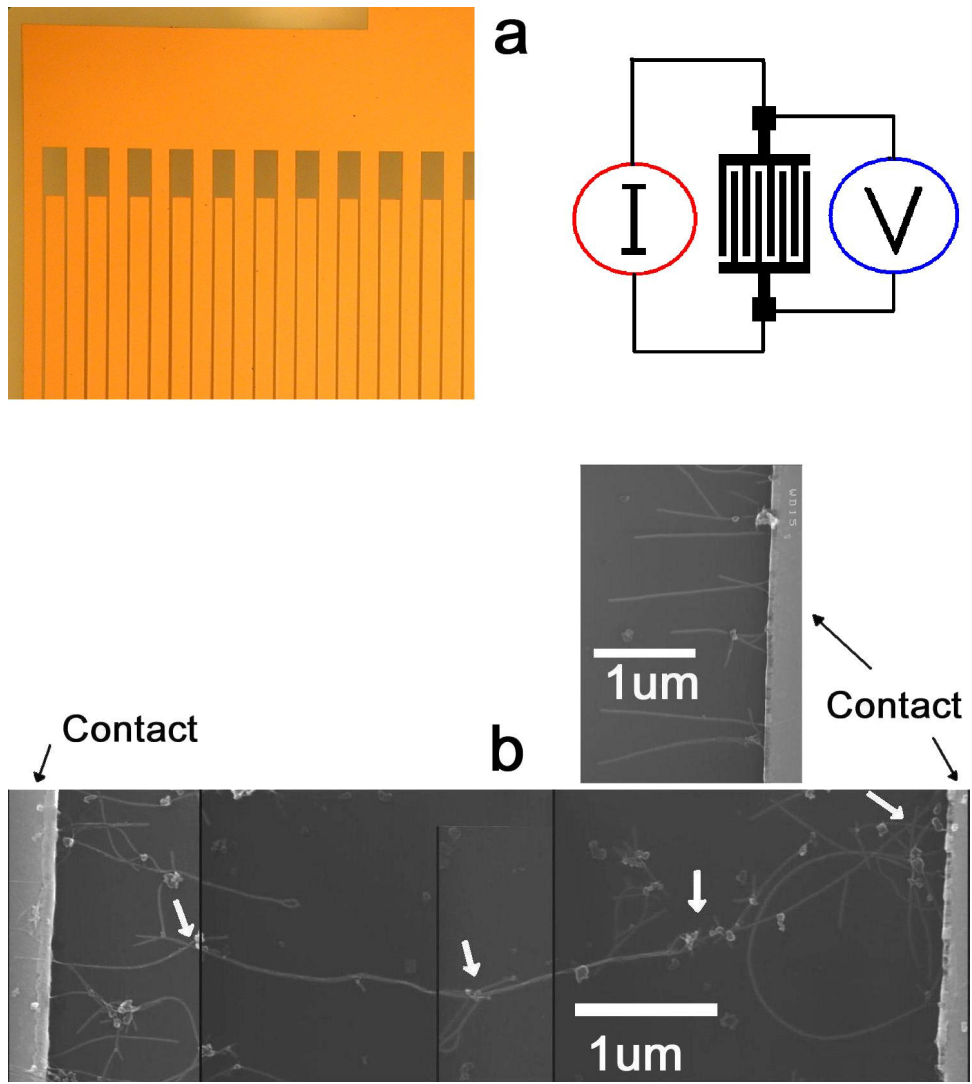

Figure 1, M. Salvato et al.

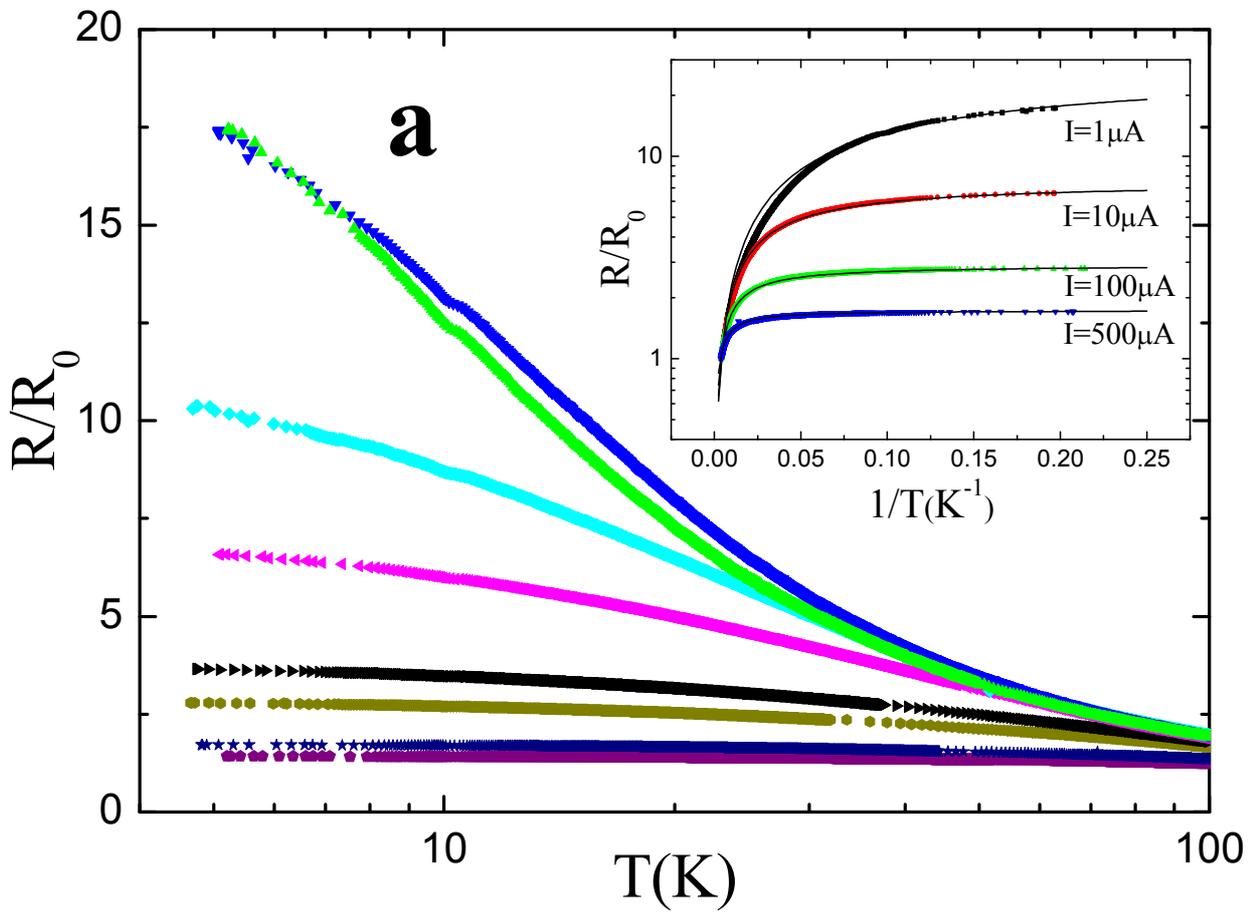

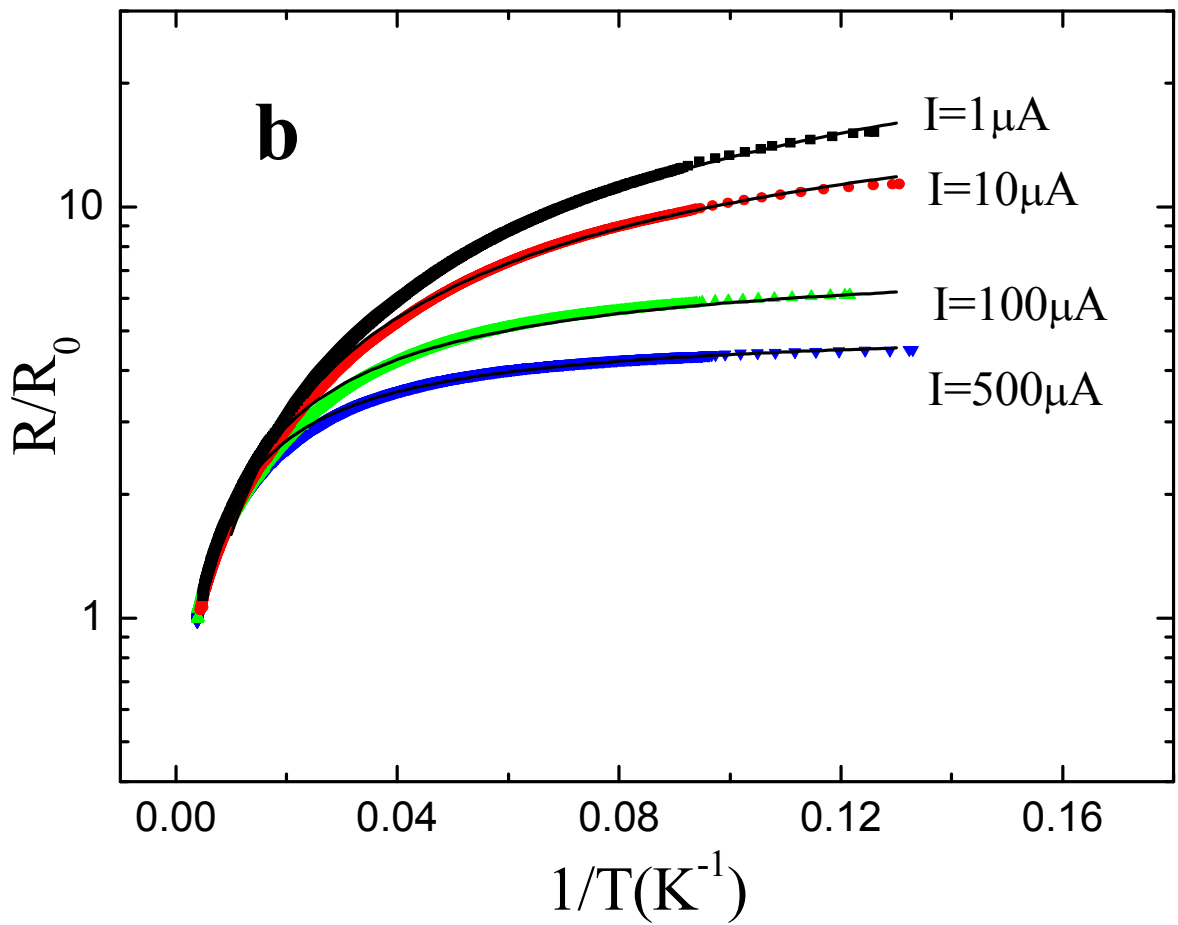

Figure 2, M. Salvato et al.

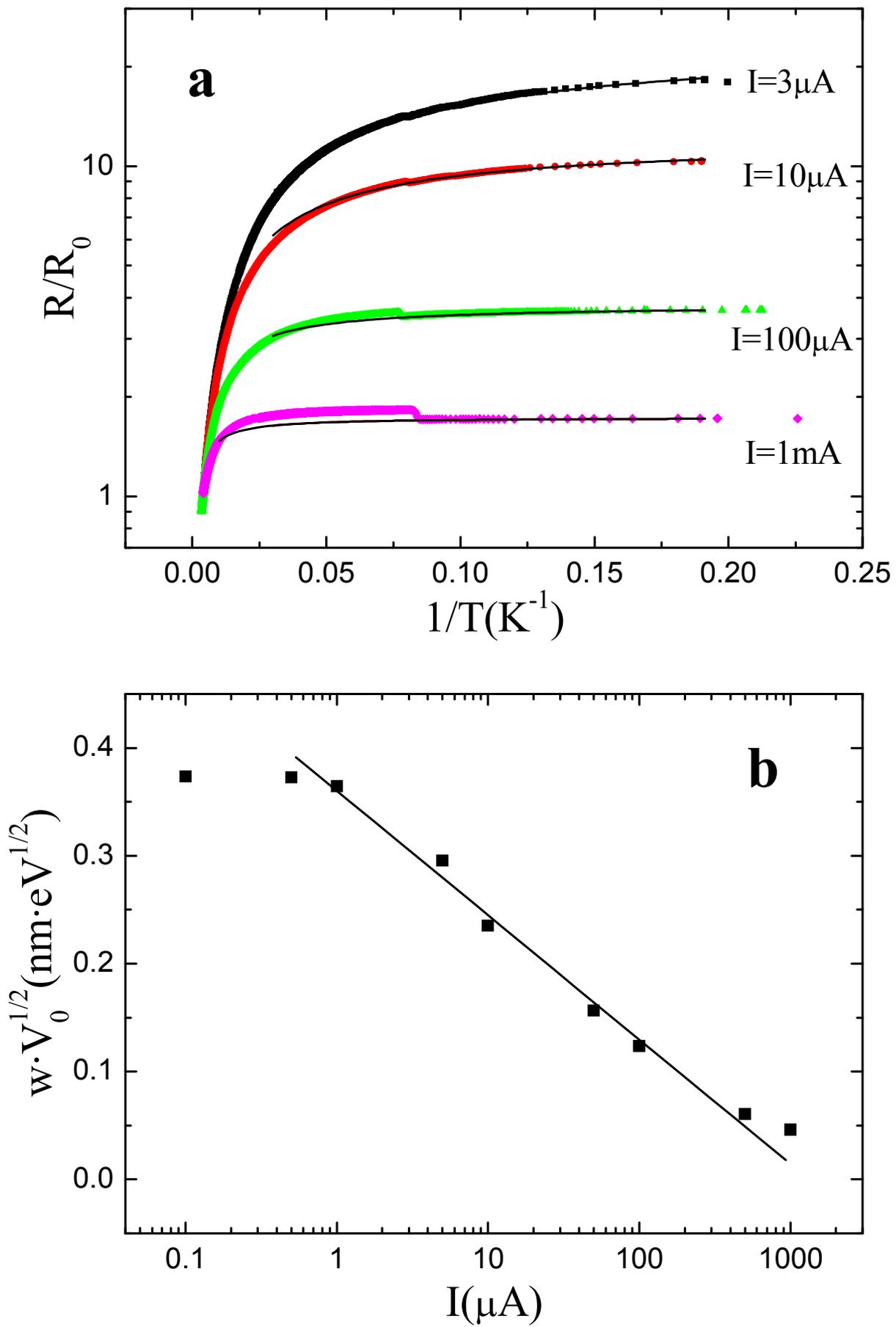

Figure 3, M. Salvato et al.

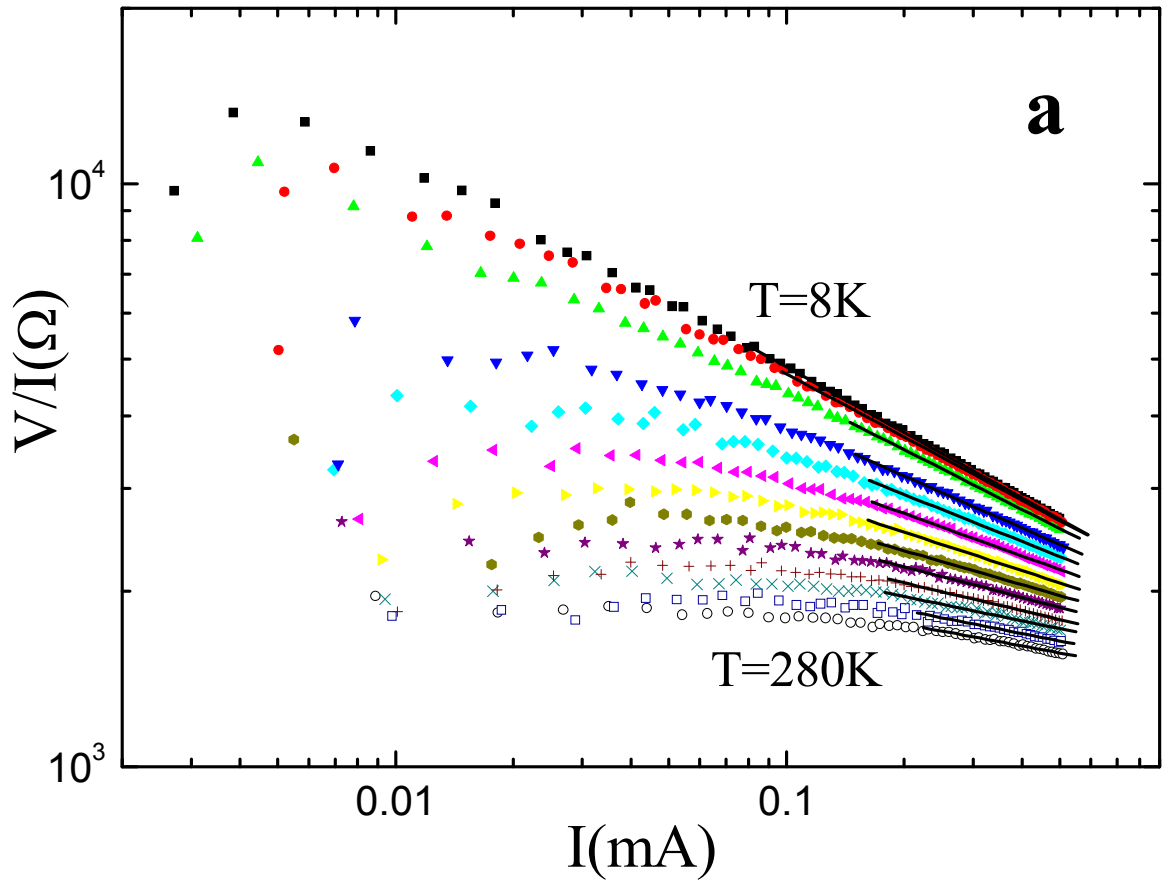
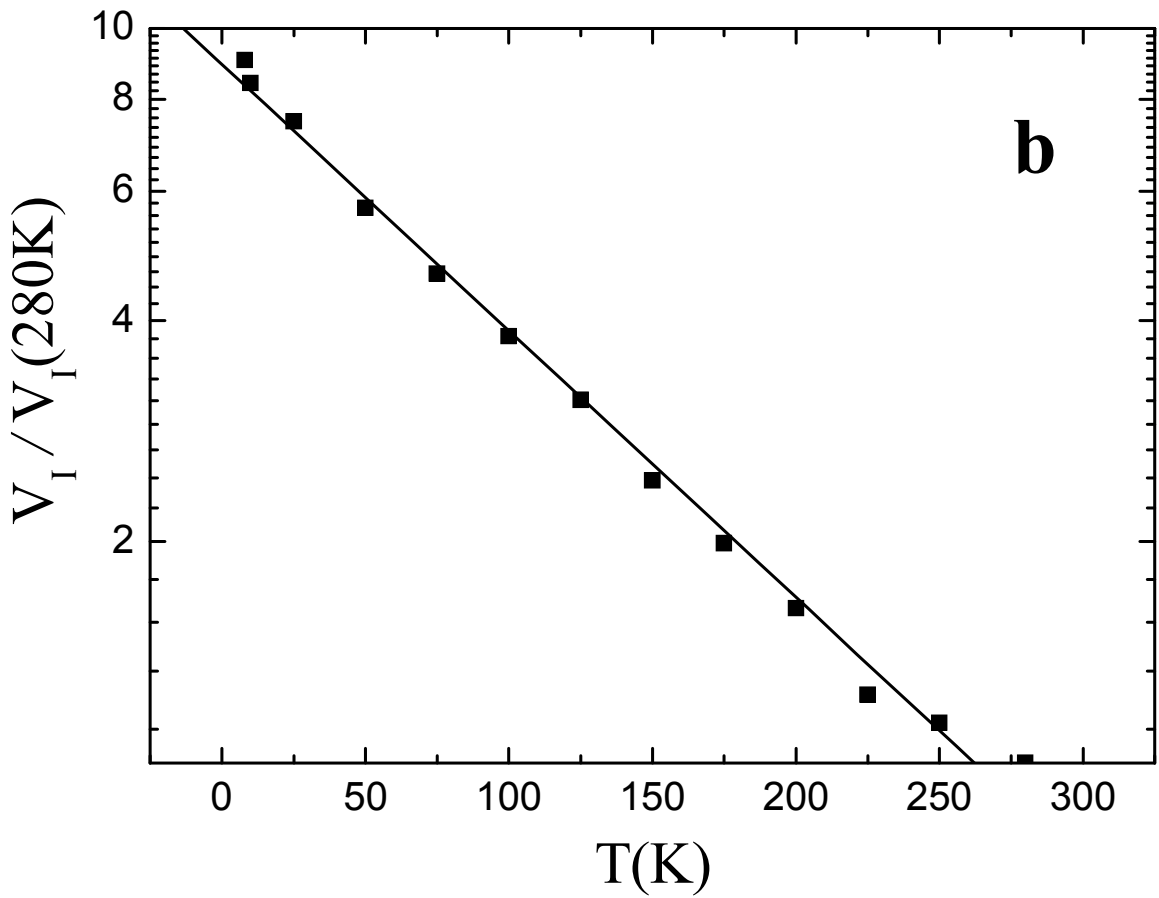

Figure 4, M. Salvato et al.